\documentclass[namedreferences]{solarphysics}
\usepackage[optionalrh,solaenum,natbib]{spr-sola-addons} % For Solar Physics 
\usepackage{graphicx}                    % For eps figures, newer & more powerfull
\usepackage{url}                         % For breaking URLs easily trough lines
\begin{document}

\begin{article}

\begin{opening}

\title{Long-term Variation of the Corona in Quiet Regions}

%%%%%%%%%%%%%%%%%%%%%%%%%%%%%%%%%%%%%%%%%%%%%%%%%%%
%% Authors Names
%
\author{S.~\surname{Kamio}$^{1}$\sep
        J.T.~\surname{Mariska}$^{2}$
       }

%%%%%%%%%%%%%%%%%%%%%%%%%%%%%%%%%%%%%%%%%%%%%%%%%%%
%% Runningheads
%
\runningauthor{S.~Kamio and J.T.~Mariska}
\runningtitle{Variation of the Corona in Quiet Regions}

%%%%%%%%%%%%%%%%%%%%%%%%%%%%%%%%%%%%%%%%%%%%%%%%%%%
%% Affiliations 
%
  \institute{$^{1}$ Max-Planck-Institut f\"ur Sonnensystemforschung,
37191 Katlenburg-Lindau, Germany\\
                     email: \url{skamio@spd.aas.org} \\ 
             $^{2}$ School of Physics, Astronomy, and Computational Sciences,
George Mason University, 4400 University Drive, Fairfax, VA 22030, USA
%                     email: \url{jtmariska@gmail.com}
             }

%%%%%%%%%%%%%%%%%%%%%%%%%%%%%%%%%%%%%%%%%%%%%%%%%%%
%%% Abstract 
\begin{abstract}
  Using {\it Hinode} EUV Imaging Spectrometer (EIS) spectra recorded daily
  at Sun center from the end of 2006 to early 2011, we studied the
  long-term evolution of the quiet corona. The light curves of the
  higher temperature emission lines exhibit larger variations in sync
  with the solar activity cycle while the cooler lines show reduced
  modulation. Our study shows that the high temperature component of
  the corona changes in quiet regions, even though the coronal
  electron density remains almost constant there.  The results suggest
  that heat input to the quiet corona varies with the solar activity
  cycle.
\end{abstract}

%%%%%%%%%%%%%%%%%%%%%%%%%%%%%%%%%%%%%%%%%%%%%%%%%%%
%% Keywords
%

\end{opening}
%-------------------------------------------------

%%%%%%%%%%%%%%%%%%%%%%%%%%%%%%%%%%%%%%%%%%%%%%%%%%%
%% Sections
%
% \section{}%\label{s:?} 

\section{Introduction}\label{s_intro}

Radiation from the corona changes dramatically over the solar activity
cycle.  \citet{peres2000} studied the long-term variation of the soft
X-ray flux measured by {\it Yohkoh}/SXT \citep{tsuneta1991} and found
that the emission measure of the whole Sun changes by orders of
magnitude between the solar activity maximum and the minimum.  The
irradiances of EUV coronal emission lines exhibit a large variation in
sync with the solar cycle, with larger amplitude at the higher
temperatures \citep{delzanna2011}.  Since the coronal irradiance is
thought to be a measure of the magnetic activity of the whole corona,
it is important to understand what features on the Sun cause the
irradiance variation.  The variation of the irradiance is mainly
attributed to magnetic field concentrations associated with active
regions, \citep[{\it e.g.},][]{foukal1990}, however, \citet{schuehle2000}
demonstrated that UV emission lines from the lower corona increased in
quiet regions during the ascending phase of the solar activity cycle.
Their results suggest that magnetic activity in quiet regions also
changes with the solar cycle.

To quantify the variation in quiet regions, coronal emission lines
formed at high temperatures need to be studied.  We investigate the
long-term variation of the EUV radiance recorded daily at Sun center
by the EUV Imaging Spectrometer \citep[EIS;][]{culhane2007} aboard
{\it Hinode} \citep{kosugi2007}.  The paper is organized as follows;
the data reduction and calibration procedure are described in Section
\ref{s_data}.  The long-term variation of emission lines and electron
density estimation are presented in Section \ref{s_results}.  The results
and interpretations are discussed in Section \ref{s_discussion}.

\section{Data Reduction}\label{s_data}

The EIS SYNOP001 and SYNOP005 studies are regular observational
programs recording EUV spectra near Sun center on a daily basis.
Since the Sun is rotating at a rate of 28 days near the equator, the
time series smoothed over the rotation period provides average
properties of the corona near Sun center.  However, one of the
challenges in determining a long-term variation of the corona is that
the sensitivity of EIS gradually decreases with time.

A detailed study of the EIS sensitivity change was carried out by
using the He~{\sc ii} $\lambda$256.32~\AA\, emission line recorded using the
regular quiet region observation studies SYNOP002 and SYNOP006 near
disk center (Mariska 2012, private communication).
Those data suggest that the degradation of the EIS
sensitivity can be expressed as
\begin{equation}
e = \{ \exp(t / \tau_1) + \exp(t / \tau_2) \} / 2,
\label{eq_deg}
\end{equation}
where $t$ denotes the time from {\it Hinode} launch on 22 September
2006, and the $e$-folding times $\tau_1$ and $\tau_2$ are 467 days and
11311 days, respectively.  In this study, the degradation of the EIS
sensitivity is compensated for using Equation (1).

A large number of individual emission line profile fits go into
each averaged data point used to compute the sensitivity correction,
and there are many possible ways to
estimate the error in each averaged data point. Two obvious
choices are to weight each averaged data point by the number of
individual measurements that went into its determination or to
weight each point by the standard deviation of all the data
points that went into the average. Each weighting results in
nearly the same results for the initial intensity and the time
constant for the shorter decay time, but different answers for
the longer decay constant. The errors for the fits, however, are
nearly the same: less than 5\% for the initial intensity, about
15\% for the shorter decay constant, and about 40\% for the
longer decay constant. The error for the longer decay constant
is larger because the value itself is large---several decades.
Thus the trend is very small over the time interval of the data
fitted in this study.

The radiance of the spectral lines listed in Table \ref{table_int} is
deduced by fitting a Gaussian function to the spectrum integrated over
100 arcsec in the north-south direction along the slit.  This integration
along the slit is intended to improve the signal to noise ratio of the
spectrum and to smooth out small-scale structures, such as coronal
bright points, along the slit.
The spatial extent in east-west direction is 1 arcsec
since the spectrum is recorded with 1 arcsec wide slit.
These emission lines exhibit
sufficient counts in quiet regions for a useful analysis.  After
correcting for the long-term degradation of the EIS sensitivity
denoted by Equation (1), data number (DN) is converted to physical units
(erg~cm$^{-2}$~s$^{-1}$~sr$^{-1}$)
by using the pre-flight calibration of the EIS effective area
\citep{lang2006,culhane2007},
which is provided in the EIS branch of Solar Software (SSW).

The light curve for each emission line is further smoothed along the
temporal axis by convolving it with a Gaussian function with a full width at
half maximum (FWHM) of 28 days---the solar rotation period.
This is intended to suppress the radiance variation caused by the
appearance and disappearance of any small active regions during
a solar rotation \citep{woods2005}.
The smoothing process at a time $t_0$ is
performed by convolving with a Gaussian weighting function
\begin{equation}
a = \exp{\left( -\frac{(t-t_0)^2}{2\sigma ^2} \right) },
\end{equation}
where $\mathrm{FWHM} = 2 \sqrt{2 \log 2}\sigma$.

\section{Results}\label{s_results}
\subsection{Location of Active Regions}\label{s_xrt}

\begin{figure} 
\centerline{\includegraphics[width=0.8\textwidth,clip=]{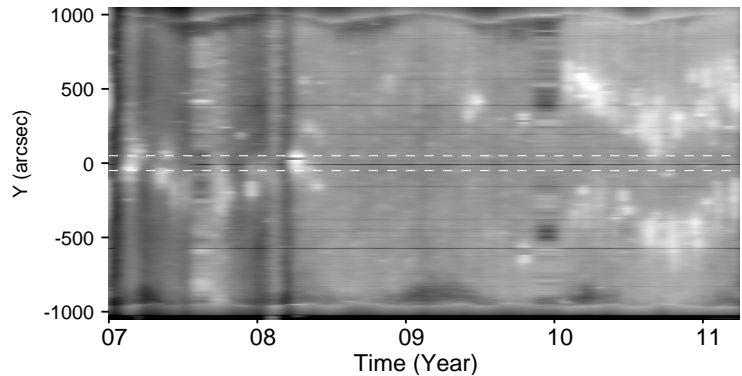}}
\caption{Time series of X-ray flux along the central meridian.
Dashed lines in the middle mark the region studied with EIS.}\label{fig_xrt}
\end{figure}

To make sure that the EIS observations at Sun center are
not directly influenced by active regions, we constructed a
time-space map by lining up stripes of central meridian
data in daily images obtained by
the X-Ray Telescope \citep[XRT;][]{golub2007} on {\it Hinode}.
Figure \ref{fig_xrt} shows a time-space map
composed of data obtained with either the Al\_poly or Al\_mesh filters,
which are available for most of the period.
Since the temperature responses of these two filters
are almost identical above 2~MK \citep{golub2007},
the appearance of active regions in these filters
should be quite similar.
The time-space map is smoothed along the temporal axis by convolving it with
the Gaussian weighting function denoted by Equation (2).
%with a 28-day width, which is the same smoothing
%process as the EIS radiance discussed in Section \ref{s_data}.
The pixel size in north-south direction remains at 2 arcsec.

The region recorded by the EIS
observations is indicated with horizontal dashed lines.
Active regions passing the meridian appear as bright features
in the plot.  Note that the EIS data are always recorded at Sun
center viewed from the Earth, and the tilt of the solar rotation axis
is not compensated for.  Active regions occasionally appeared near Sun
center from 2007 to the middle of 2008, while they are almost absent
in 2009.  In 2010, they again started to appear regularly, but distant
from Sun center.  The plot demonstrates that with only one
exception in 2008 the EIS observing region is not covered by active
regions. Therefore, the radiances deduced from the EIS observations
are not directly affected by an increase or a decrease in the number
of active regions.

\subsection{EUV Radiance Variation}\label{s_radiance}

\begin{table}
\begin{tabular}{cccccc}
\hline
Ion & $\lambda$ &  $\log T_{\mathrm{e}}$ & Dec 2006 & Feb 2009 & Ratio \\
 & [\AA] & [K] &  \multicolumn{2}{c}{[erg cm$^{-2}$ s$^{-1}$ sr$^{-1}$]} & \\
\hline
Fe~{\sc xiv} \tabnote{With a blending Fe~XI 264.77\AA} & 264.79 & 6.25 & 38 & 7.8 & 4.8 \\
Fe~{\sc xiii} & 202.04 & 6.20 & 207 & 33 & 6.2 \\
Fe~{\sc xii}  & 193.51 & 6.15 & 225 & 70 & 3.2 \\
Si~{\sc x}    & 258.37 & 6.15 & 75 & 35 & 2.1 \\
Fe~{\sc xi}   & 180.40 & 6.10 & 426 & 282 & 1.5 \\
Fe~{\sc x}    & 184.54 & 6.00 & 103  & 118 & 0.86 \\
Si~{\sc vii}  & 275.36 & 5.80 & 16  & 23 & 0.71 \\
Fe~{\sc viii} & 185.21 & 5.65 & 31  & 44 & 0.72 \\
He~{\sc ii}   & 256.32 & 4.85 & 256 & 229 & 1.1 \\
\hline
\end{tabular}
\caption{Fitted radiance at the beginning of {\it Hinode} observations (December 2006) and at the minimum (February 2009).}\label{table_int}
\end{table}

\begin{figure} 
\centerline{\includegraphics[width=\textwidth,clip=]{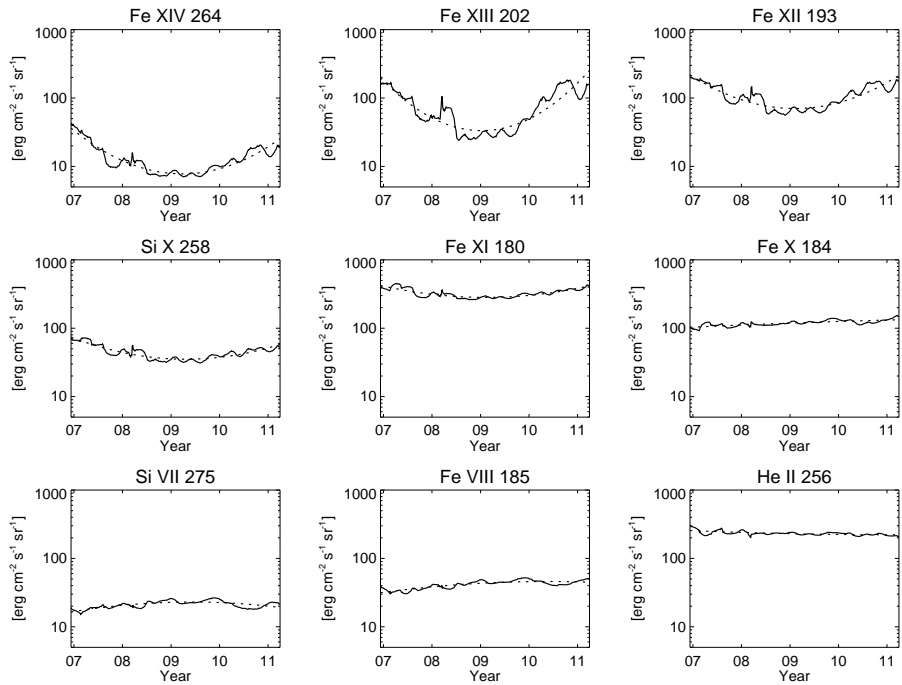}}
\caption{Radiances measured in different emission lines are plotted as
  solid curves.  The dotted curve in each panel show a second-order
  polynomial fit to the measured radiance.}\label{fig_lc}
\end{figure}

Figure \ref{fig_lc} shows the smoothed light curve in each emission
line.
The measured radiance is smoothed by convolving with a Gaussian
function with a width of the 28-day solar rotation period
(see Section \ref{s_data}).
Since the
light curves are nearly symmetric with respect to the minimum in 2009,
the smoothed radiance is fitted with a second-order polynomial
function to determine the long-term variation.  The fitted polynomial
functions are overplotted using dotted curves in Figure \ref{fig_lc},
and show a reasonable agreement with the light curves.  In the
following, the radiance in each emission line is determined by using
the fitted polynomial functions.  The lowest radiance for Fe~{\sc xii} is
attained in February 2009, which is regarded as the minimum of the
long-term variation.  (The NOAA Space Weather Prediction Center lists
the solar cycle minimum date as December 2008.)  The radiance values
at the beginning of {\it Hinode} observations (December 2006)
and at the minimum (February 2009) are summarized in Table
\ref{table_int}.

Emission lines hotter than Fe~{\sc x} tend to show larger variation ratios
with increasing temperature. The only exception is the Fe~{\sc xiv} line,
which has a smaller ratio than that of the Fe~{\sc xiii} line.  A possible
effect of a blending emission line is discussed in Section \ref{s_dem}.  On
the contrary, the low-temperature emission lines between Fe~{\sc viii} and
Fe~{\sc x} show constant or a slightly larger radiance at the minimum.
Although the sensitivity change is determined from the He~{\sc ii} emission
line alone, a wavelength dependent degradation of the instrument is
not likely the reason for this trend, because Fe~{\sc viii} and Si~{\sc vii}
are observed in different EIS wavelength channels and yet exhibit a
similar variation (Table \ref{table_int}).

\subsection{Spatial Distribution of the Radiance}\label{s_histogram}
\begin{figure} 
\centerline{\includegraphics[width=0.5\textwidth,clip=]{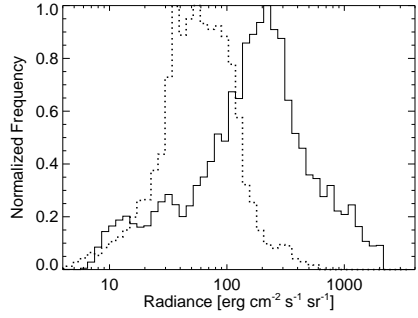}
}
\caption{Histogram of the Fe~{\sc xii} $\lambda$193.51\AA\, radiance.
Solid and dotted lines show the distributions
in the December 2006 data and in the February 2009 data, respectively.}\label{fig_histogram}
\end{figure}

To study the spatial distribution of the radiance,
Figure \ref{fig_histogram} compares the histogram of the Fe~{\sc xii}
$\lambda$193.51\AA\, radiance in each pixel of EIS
recorded in December 2006 and in February 2009.
The intense Fe~{\sc xii} $\lambda$193.51\AA\ emission line
allows us to analyze the radiance in an individual pixel (1''$\times$1'')
without summing,
which can differentiate between localized and dispersed brightenings.
The histogram is peaked at
200~erg~cm$^{-2}$~s$^{-1}$~sr$^{-1}$ in the December 2006 data (solid line),
while that in February 2009 is at 60~erg~cm$^{-2}$~s$^{-1}$~sr$^{-1}$
(dotted line).  The ratio of these values is close to the mean
radiance ratio of 3.2 in Table \ref{table_int}.  The histogram
demonstrates that the long-term variation of the average radiance is
not caused by an increase or a decrease of localized bright features,
but instead by a variation of the entire region.

\subsection{Electron Density}\label{s_density}
\begin{figure} 
\centerline{\includegraphics[width=0.5\textwidth,clip=]{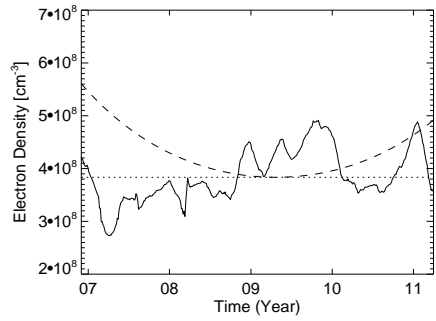}}
\caption{Time history of the electron density derived from the Si~{\sc x} line pair.}\label{fig_density}
\end{figure}

The electron density is estimated from the line ratio of Si~{\sc x}
$\lambda$258.37~\AA /261.04~\AA.
The line ratio is deduced from the Si~{\sc x} radiance
integrated over 100 arcsec along the slit.
The relationship between electron
density and the line ratio is calculated using version 7.1 of the
CHIANTI atomic database \citep{dere1997,landi2012}.
Figure \ref{fig_density} presents the electron density variation.  The
obtained electron density is smoothed along the temporal axis in the
same manner as the radiance in Section \ref{s_radiance}.  The dotted line
plotted over Figure \ref{fig_density} indicates a constant density of
$3.8\times 10^8$~cm$^{-3}$.  In contrast to the radiance variation in
the Si~{\sc x} emission line, the electron density shows no clear trend of
solar cycle variation.
The uncertainly of the radiance due to the spectrum fitting
 is estimated to be smaller than 10\%,
which corresponds to an uncertainty of 30\% in the density estimation.
Note that the line ratio analysis depends solely on the relative radiance
and is not affected by the uncertainty of the absolute radiance.

In the case of an isothermal plasma, the observed radiance $I$ is denoted
as
\begin{equation}
I = f d n_{\mathrm{e}}^2 g(T,n_{\mathrm{e}})
\label{eq_em}
\end{equation}
where $f$, $d$, and $g(T,n_{\mathrm{e}})$ are the filling factor of the emitting
plasma, the depth over which the emission line is formed, and the
contribution function to the line, respectively.  To examine whether
the density variation is responsible for the radiance variation, a
square root of the Si~{\sc x} $\lambda$258.37\AA\, radiance is overplotted
in Figure \ref{fig_density} with a dashed curve.  If the density
variation is the primary cause of radiance variation, the density
should agree with the square root of radiance.
Figure \ref{fig_density} indicates that the density variation is
not sufficient to reproduce the observed radiance variation.
Other factors, the filling
factor $f$, depth of the corona $d$, or the contribution function, are
thus responsible for the radiance variation.

\subsection{DEM Analysis}\label{s_dem}
\begin{figure} 
\centerline{\includegraphics[width=0.5\textwidth,clip=]{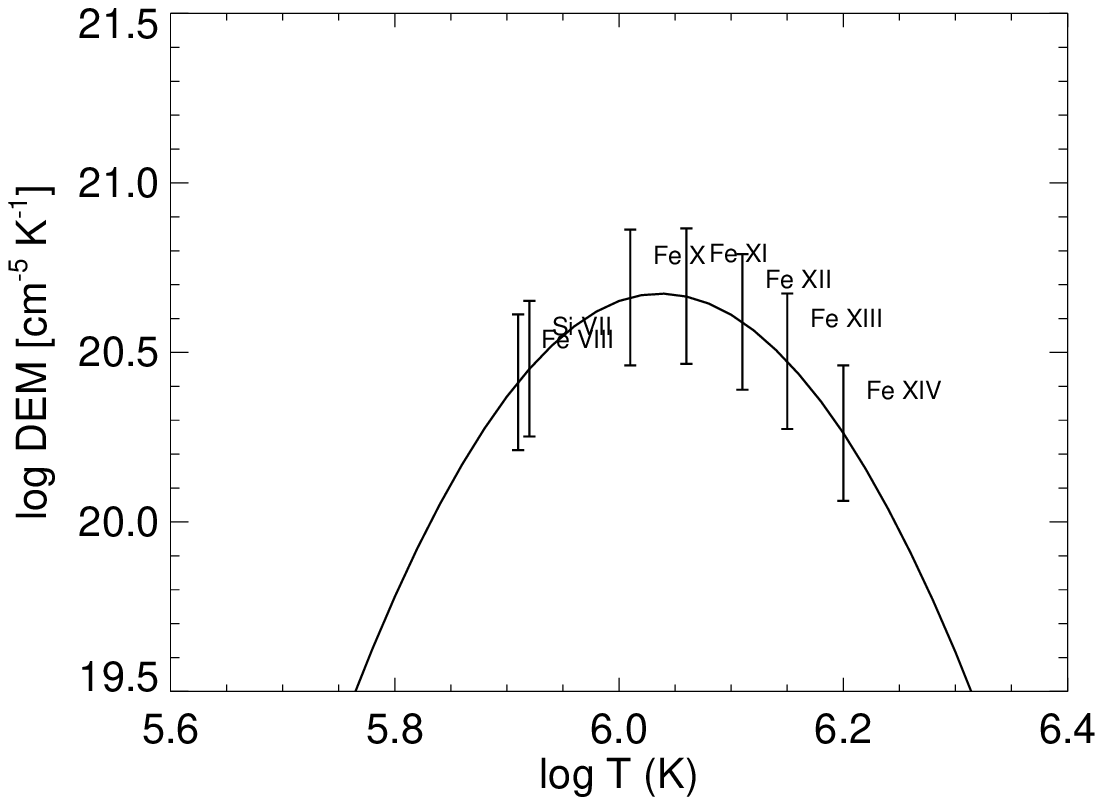}
\includegraphics[width=0.5\textwidth,clip=]{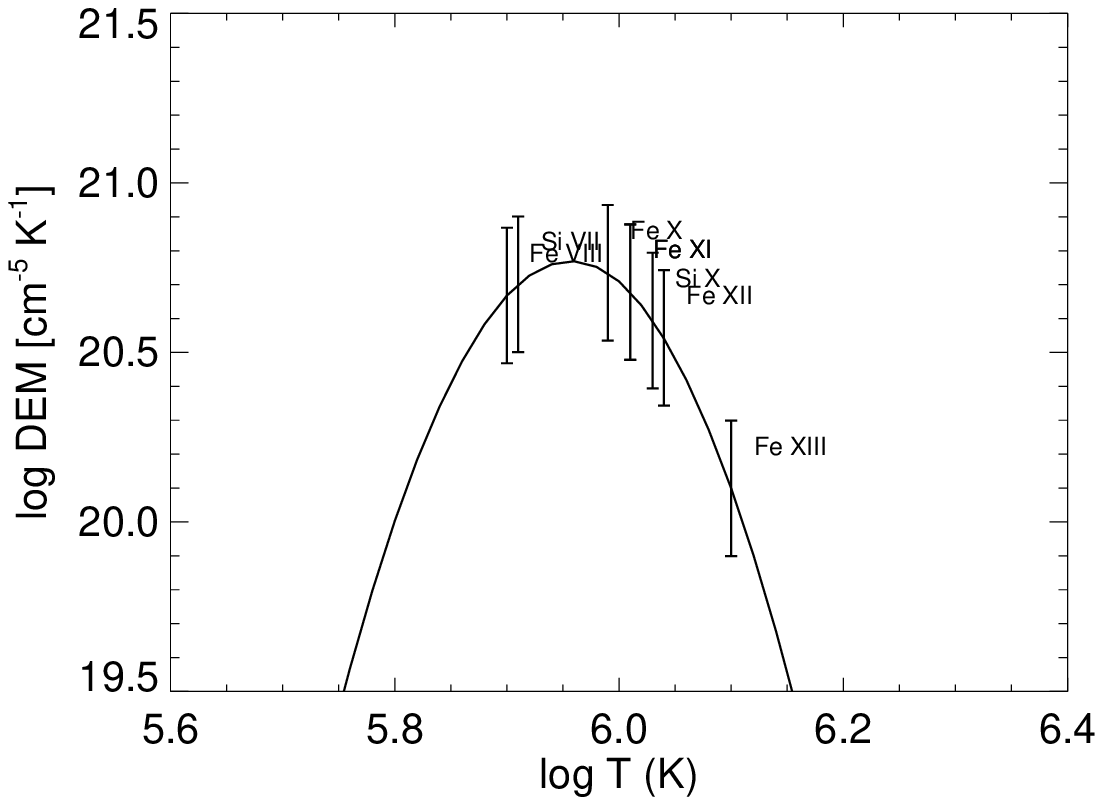}}
\caption{Results from DEM analysis in December 2006 (left panel) and February 2009 (right panel).}\label{fig_dem}
\end{figure}

A differential emission measure (DEM) analysis is carried out with the
aid of the CHIANTI atomic database.  All emission lines in Table
\ref{table_int} except for He~{\sc ii} are used in the analysis. Since a
sufficient number of emission lines are available only within a
temperature range of $\log T_{\mathrm{e}}$ = 5.65--6.25, we restrict
the analysis to that interval.  A DEM in that
temperature range is computed by using the procedure provided in the
CHIANTI atomic database.  It deduces an optimum DEM so that the DEM
multiplied by the contribution function of each emission line
reproduces the measured radiance.  Figure \ref{fig_dem} presents the
DEM at two different times.  In December 2006, the DEM peak is located
at $\log T_{\mathrm{e}}$ = 6.05, while that at the minimum in February 2009 the peak is
found at 5.95.
The high
temperature tail of the DEM is extended higher in the former period
than the minimum period.  This explains the trend that the higher
temperature emission lines exhibit greater decreases in
the minimum period (Figure \ref{fig_lc}).
In contrast, the lower temperature emission lines
(Fe~{\sc viii} and Si~{\sc vii}) indicate slight increases
in the minimum period
as the DEM peak is shifted towards the lower temperature.
These results show that the properties of the quiet region corona
change with the solar activity cycle.

The radiance in the Fe~{\sc xiv} $\lambda$264.79~\AA\ line is a special
case. It turns out that the radiance of Fe~{\sc xiv} $\lambda$264.79~\AA\ is
overtaken by a blending Fe~{\sc xi} $\lambda$264.77~\AA\ feature during the
minimum period.  According to the CHIANTI estimation of the radiance,
Fe~{\sc xiv} is five times larger than Fe~{\sc xi} in December 2006,
while Fe~{\sc xiv}
gets smaller than Fe~{\sc xi} at the minimum in February 2009. 
Because of
the Fe~{\sc xi} $\lambda$264.77~\AA\ blending, the measured radiance at
$\lambda$264.79~\AA\ (Table \ref{table_int}) does not decrease much,
resulting in a smaller variation than the Fe~{\sc xiii} $\lambda$202.04~\AA\
light curve.  Thus care should be taken in analyzing the Fe~{\sc xiv}
$\lambda$264.79~\AA\ radiance.

\section{Discussion}\label{s_discussion}

We found a long-term variation of the EUV emission lines in quiet
regions in sync with the solar activity cycle.  The time series of the
EIS spectra was recorded at Sun center, where there is minimal
contamination from active regions (Figure \ref{fig_xrt}).  The
light curves of emission lines above 1~MK clearly show that the quiet
region corona varies with the solar activity cycle
(Figure \ref{fig_lc}).  The radiance variation is attributed to the
coronal temperature rather than the coronal electron density.  As
shown in Figure \ref{fig_density}, the observed density variation is inadequate
to reproduce the observed radiance variation.  The temperature
dependent variation of the light curves is explained as a modulation of
the high temperature component in the corona (Figure \ref{fig_dem}).

\citet{orlando2001} claimed that the average temperature of low X-ray
flux regions observed by the {\it Yohkoh}/SXT remained constant around
1.5~MK between the solar maximum and the minimum.  A possible
explanation for this disagreement is that the sensitivity of SXT is
low below 1~MK, and hence is not optimum for diagnosing the corona
around 1~MK.  The EIS observations provided us with several emission
lines at that temperature to allow a more detailed study.

Our study proves that the corona in quiet regions is not constant, but
changes with the solar activity cycle.  It suggests that the heat
input to the quiet corona changes with the solar activity cycle.  In
addition, the spatial distribution of the radiance indicates that the
radiance variation in the quiet region is a dispersed phenomenon
rather than localized features (Figure \ref{fig_histogram}).
Since a nearly linear relationship has been established
between total magnetic flux and X-ray radiance in quiet regions,
active regions, and stars \citep{pevtsov2003},
our finding of coronal radiance variation is
attributed to a change in total magnetic flux in the quiet region.
\citet{close2003} argued that quiet regions are filled with short
loops connecting magnetic flux fragments in the photosphere, forming
the so-called magnetic carpet.
We speculate that the total magnetic flux of the
small-scale magnetic loops in the corona changes with the solar activity cycle, and 
hence the heat input to the corona.
Indeed, \citet{pauluhn2003} studied magnetograms obtained in the quiet
regions by SOHO/MDI and found that the fraction of network area
increased from 10\% near the solar activity minimum to 13\% near the
maximum.
The long-term variation of magnetic fields in the corona
needs to be studied in order to understand the coronal heating process
in quiet regions.  Our results should encourage modellers to consider how the
small-scale magnetic fields in quiet regions are related to the solar
activity cycle on a global scale.  It must be noted that our
observations covers only five years around the solar minimum.
Therefore, the variation amplitude along the solar cycle remains to be
studied when the solar maximum is reached.

%%%%%%%%%%%%%%%%%%%%%%%%%%%%%%%%%%%%%%%%%%%%%%%%%%%%%%%%%%%%%%%%%%%%%%%%%%%
%% Acknowledgements
%
\begin{acks} {\it Hinode} is a Japanese mission developed and launched
  by ISAS/JAXA, with NAOJ as domestic partner and NASA and STFC (UK)
  as international partners.  It is operated by these agencies in
  co-operation with ESA and NSC (Norway). JTM acknowledges support from
  the NASA {\it Hinode} contract with the Naval Research Laboratory.
\end{acks}

%%% %%%%%%%%%%%%%%%%%%%%%%%%%%%%%%%%%%%%%%%%%%%%%%%%%%%%%%%%%%%
%% Bibliography
%
% Using BibTeX
%
\bibliographystyle{spr-mp-sola-cnd}
% %\bibliographystyle{spr-mp-sola-cnd} %% Alternative style: no title, no concluding page
\bibliography{reference}  
%
% Without BibTeX 
% \begin{thebibliography}{}
% \bibitem[\protect\citeauthoryear{Author}{Year}]{key}
%   <bibliographical entry>
%
% \bibitem[\protect\citeauthoryear{}{}]{}
%   
%  
% \end{thebibliography}

\end{article} 
\end{document}